# PICTS: A Novel Deep Reinforcement Learning Approach for Dynamic P-I Control in Scanning Probe Microscopy

*Ziwei Wei,*[+,1] *Shuming Wei,*[+,1] *Qibin Zeng,*[2] *Wanheng Lu,*[3] *Huajun Liu,*[2] *and Kaiyang Zeng**[1,4]

1. Department of Mechanical Engineering, National University of Singapore, 9 Engineering Drive 1, Singapore, 117576.
2. Institute of Materials Research and Engineering, Agencies of Science, Technology and Research (A*STAR), 2 Fusionopolis Way, Innovis, #08-03, Singapore, 138634.
3. Department of Electrical and Computer Engineering, National University of Singapore, 4 Engineering Drive 3, Singapore 117583.
4. NUS Research Institute (NUSRI), No. 377 Linquan Street, Suzhou Industrial Park, Suzhou, Jiangsu Province, China, 215123.

+ Equal contributions
* Corresponding author (mpezk@nus.edu.sg)

## Abstract

Scanning Probe Microscopy (SPM) often faces great challenges when dealing with nonlinear and time-varying behaviours, especially during scanning the samples with complex structures/properties or abrupt topographical changes. Traditional controller with fixed proportional-integral (P-I) parameters usually face challenges to maintain stability and precision under these conditions, leading to substantial artefacts, errors and instability, specially, the materials with sharp edges, multiphase soft structures, or significant topographical variations can disrupt probe-sample interactions, resulting in nonlinear responses that are difficult to manage with conventional control strategies. To address these challenges, in this study, we develop a new control system, named Parallel Integrated Control and Training System (PICTS), which leverages deep reinforcement learning (DRL) to train an agent capable of monitoring the system state in real-time and dynamically adjusting the control strategy. This adaptive approach enables precise control, even when dealing with sudden changes in sample properties and/or topographical features. Experimental results demonstrate that the DRL controller significantly outperforms that of the commercial controller with fixed P-I parameters, reducing deflection errors by 26% to 90% across various samples and conditions. Statistical analyses show a higher concentration of error values around zero and fewer outliers, indicating improved stability and scanning precision. Moreover, the PICTS architecture efficiently balances real-time control and computational demands by leveraging a field-programmable gate array (FPGA) to handle critical tasks, allowing it to operate effectively without needing high-performance computers. By integrating DRL into SPM control systems, this work offers a robust solution for complex and dynamic scanning environments, enhancing image quality and expanding the capabilities of SPM technology. This approach also demonstrates the potential of advanced machine learning techniques to overcome the limitations of traditional control methods, paving the way for further innovations in the field.

## 1. Introduction

Scanning Probe Microscopy (SPM) stands as a cornerstone technology in nanoscale science, offering unparalleled capabilities for imaging and manipulating materials with atomic-level precision.[1–4] Its versatility has made it indispensable across multidisciplinary fields such as materials science,[3] physics,[5] biology,[1,2] and chemistry.[6] SPM techniques enable researchers to measure and characterise various materials, from soft biological tissues to hard materials, insulating/conductive surfaces, also under various environmental conditions, including air, liquid, and vacuum.[1,3–6] By physically interacting with the sample surface using a sharp probe, SPM achieves high-resolution images of the surface and functional mapping of mechanical, electrical, magnetic, chemical properties and many others.

The accuracy and reliability of SPM measurements heavily depend on the precision of the control systems, particularly the height control loop that governs the vertical movement of the Z-actuator.[7,8] Unlike optical or electronic microscopy, SPM's reliance on direct interaction between the probe and sample makes it highly sensitive to instabilities in the feedback control system. Any oscillations or delays in the Z-actuator control loop can lead to distorted topographical images, artefacts, loss of contact with the sample surface, and compromised functional measurements.[1,8,9] These issues significantly affect height accuracy and obscure critical differential signals essential for advanced SPM-based techniques. Some examples of artefacts can be found in Section S1 and Figure S1 of the Supplementary Information (SI).

Furthermore, numbers of advanced SPM-based techniques, such as Kelvin Probe Force Microscopy (KPFM), Piezoresponse Force Microscopy (PFM), and Conductive-AFM (C-AFM), require the Z-actuator to maintain a precise position at/from the sample surface to detect subtle, real-time changes in the probe–sample interactions.[10–13] For instance, in KPFM, the constant distance between the tip and samples is used to map the electrostatic force accurately. Instabilities in the height control loop can severely affect these measurements, leading to data distortion and undermining the integrity of quantitative analyses across mechanical, thermal, and electrical properties.

Traditionally, SPM systems have relied on Proportional-Integral (P-I) controllers, which generally perform well in linear and time-invariant systems by inputting fixed P-I parameters. However, due to their inherent limitation, i.e., lacking direct knowledge of the processes, P-I controllers must act as passive “observers” when the precise mathematical model is unavailable. This results in reactive, often compromised control responses rather than optimal adjustments, which becomes evident in the highly nonlinear and dynamic environment of the SPM. Each probe-sample combination requires unique control mechanism, especially when the controller faces unpredictable probe degradation, as well as sharp edges, heterogeneous structures, or significant topographical variations at the sample surface. In such cases, controller with fixed P-I parameters struggles in response to rapid changes in the surface and/or properties, lacking the flexibility to maintain effective control, which results in lagging control, errors and inconsistent performance. Hence, more adaptive and responsive control strategies are critical to ensure consistent and reliable performance as SPM applications expand across a diverse range of materials/systems and probes.

The rise of artificial intelligence (AI) offers promising avenues for developing such adaptive "observers". Machine learning techniques, notably neural networks and reinforcement learning[14] have shown potential for automating and enhancing SPM control systems.[15,16] For example, neural networks excel in learning complex non-linear relationships and have been employed to conduct automated experiments, acquire high-resolution images without user intervention,[17–19] evaluate tip degradation,[20] select optimal scanning parameters[16] and enhance control performance.[21] However, neural network approaches typically rely on offline training, which restricts their adaptability to real-time, dynamic system changes. Once trained, their behaviour is static, which may result in suboptimal performance under evolving or unforeseen conditions. Reinforcement learning, particularly in Markov Decision Processes (MDPs), offers a more dynamic solution by enabling continuous learning through interaction with the environment. This allows algorithms to adjust the scanners in real time, making them more responsive to the variations typically encountered in SPM operations. However, direct reinforcement learning control without a P-I intermediary can lead to non-ideal outcomes, such as jagged oscillations, due to the discrete decision-making nature of the MDPs, which is undesirable in SPM control, where smoothness and stability are crucial.[21] While reinforcement learning does offer some methods for handling continuous action spaces, their performance is often limited by insufficient function approximation and low exploration efficiency in high-dimensional spaces.

Deep reinforcement learning (DRL) algorithms combine the strengths of both neural networks and reinforcement learning while mitigating their limitations.[22] DRL has been explored in SPM for various scenarios, such as finding optimal SPM probe trajectories,[23] avoiding hard contact between the probe and sample,[24] and improving tracking. However, very few works have been undertaken to directly enhance the control performance of the SPM controller. While some research has demonstrated in simulations that DRL has the potential to optimise SPM control systems,[24,25] several issues persist:

a) To our knowledge, no prior work has successfully integrated DRL into a commercial SPM control system to improve performance, nor has DRL's adaptability been fully utilised in practical settings.
b) Testing and training environments in existing studies are often simplified to repetitive patterns, such as sine or square waveforms. These sidestep two primary challenges in the SPM control: i.e., achieving high robustness in continuously changing and unknown environments as well as balancing immediate performance with long-term stability.
c) High-performance computing methods demand computational resources beyond the standard hardware typically available in laboratories with the commercial SPM systems.

To bridge this gap, in this work, we developed a new control system which is named the "Parallel Integrated Control and Training System" (short as PICTS); this system integrates the state-of-the-art deep reinforcement learning algorithm, Soft Actor-Critic (SAC) with Automating Entropy Adjustment,[26] into tuning the commercial SPM controller with fixed P-I parameters. SAC offers several advantages, including efficient exploration in continuous action spaces and optimising cumulative rewards, allowing it to discover globally optimal solutions while dynamically adapting to environmental changes.[26,27] This balance is essential for SPM operations, where specific parameter settings may deliver strong short-term results but compromise overall system stability. Hence, a necessary feature of the PICTS is its ability to balance real-time control requirements with the computational demands. The system maintains high-frequency responsiveness without overloading standard computational resources by using a field-programmable gate array (FPGA) for critical tasks and offloading computationally intensive processes. This design ensures that advanced control capabilities are accessible to the users without requiring high-end computational setups, making the technologically practical and scalable.

The core characteristics and contributions of this study can be summarised as follows:

a) This work demonstrates the synergy between feedback control systems and DRL. In SPM, the system dynamics are often not fully understood or are difficult to model precisely. DRL provides a framework for the controller to learn optimal behaviours through trial-and-error interactions with the environment, allowing the controller to adapt to the highly nonlinear and time-varying conditions inherent in the SPM, hence to overcome the limitations of commercial controllers with fixed P-I parameters.
b) With the design and implementation of the PICTS, it is demonstrated that PICTS has the strong ability to enhance precision and stability in the real-time SPM operations, particularly for the samples with challenging features such as sharp edges, multiphase soft structures, and significant topographical variations.
c) The experimental results show that the DRL controller significantly outperforms that of the commercial controllers, reducing deflection errors by 26% to 90% across multiple samples and conditions. The error distributions also improve significantly, with the DRL controller exhibiting fewer outliers and a higher concentration of error values near zero.

Overall, integrating DRL into SPM control systems offers a robust solution for complex and dynamic scanning environments, enhancing image quality and expanding the capabilities of SPM technology. This approach demonstrates the practical feasibility of applying advanced machine learning techniques in the SPM operations, paving the way for future innovations of SPM control and imaging.

## 2. Results

The following sections provide a detailed analysis of how the PICTS achieves superior control performance (Section 2.1). Section 2.2 presents experimental results from the commercial SPM system with the new controller designed in this work, demonstrating that the PICTS delivers improved measurement accuracy compared with that from the commercial SPM control systems, particularly when scanning challenging samples.

### 2.1 Parallel Integrated Control and Training System (PICTS)

In this study, PICTS integrates high-speed control with adaptive deep reinforcement learning capabilities for SPM controlling and imaging. In this section, we first present the system design, key innovations, and the role of the reward function in the PICTS, highlighting how PICTS can enhance SPM control and performance.

#### *2.1.1 System Overview*

In this study, PICTS is designed to integrate seamlessly with a commercial SPM system (MFP-3D, Asylum Research, Oxford Instruments, CA, USA) while operating independently on its dedicated host computer. The "Host Computer" defined here, is part of the PICTS, which manages agent decision-making and training, whereas the commercial SPM system operates by its own computer to perform physical scanning tasks. The system achieves coordination through three operational cycles that balance the trade-off between the real-time control requirements and computational efficiency (Figure 1):

Cycle 1: Real-Time Control Loop on FPGA
Cycle 2: Agent Decision-Making Loop on Host Computer
Cycle 3: Training Loop on Host Computer

This multi-cycle architecture allows PICTS to achieve high-frequency control while continuously adapting to the dynamic environment of the SPM operations.

### *2.1.2 Innovative Architecture and Data Flow*

Figure 1 illustrates the system architecture and data flow with the PICTS. The data flow initiates when the SPM tip approaches the setpoint, triggering the initial Proportional-Integral (P-I) parameters on the FPGA-based controller. The controller's output adjusts the Z-axis actuator of the SPM system via a Digital-to-Analogue Converter (DAC). Upon tip-sample contact, a deflection voltage is generated, captured by an Analogue-to-Digital Converter (ADC), and processed by the FPGA to calculate the error signal, i.e., the difference between the deflection voltage and the setpoint. The host PC reads the updated system state, determining the following policy and action. The adjusted P-I gains are then applied to the commercial P-I controller, completing a learning cycle.

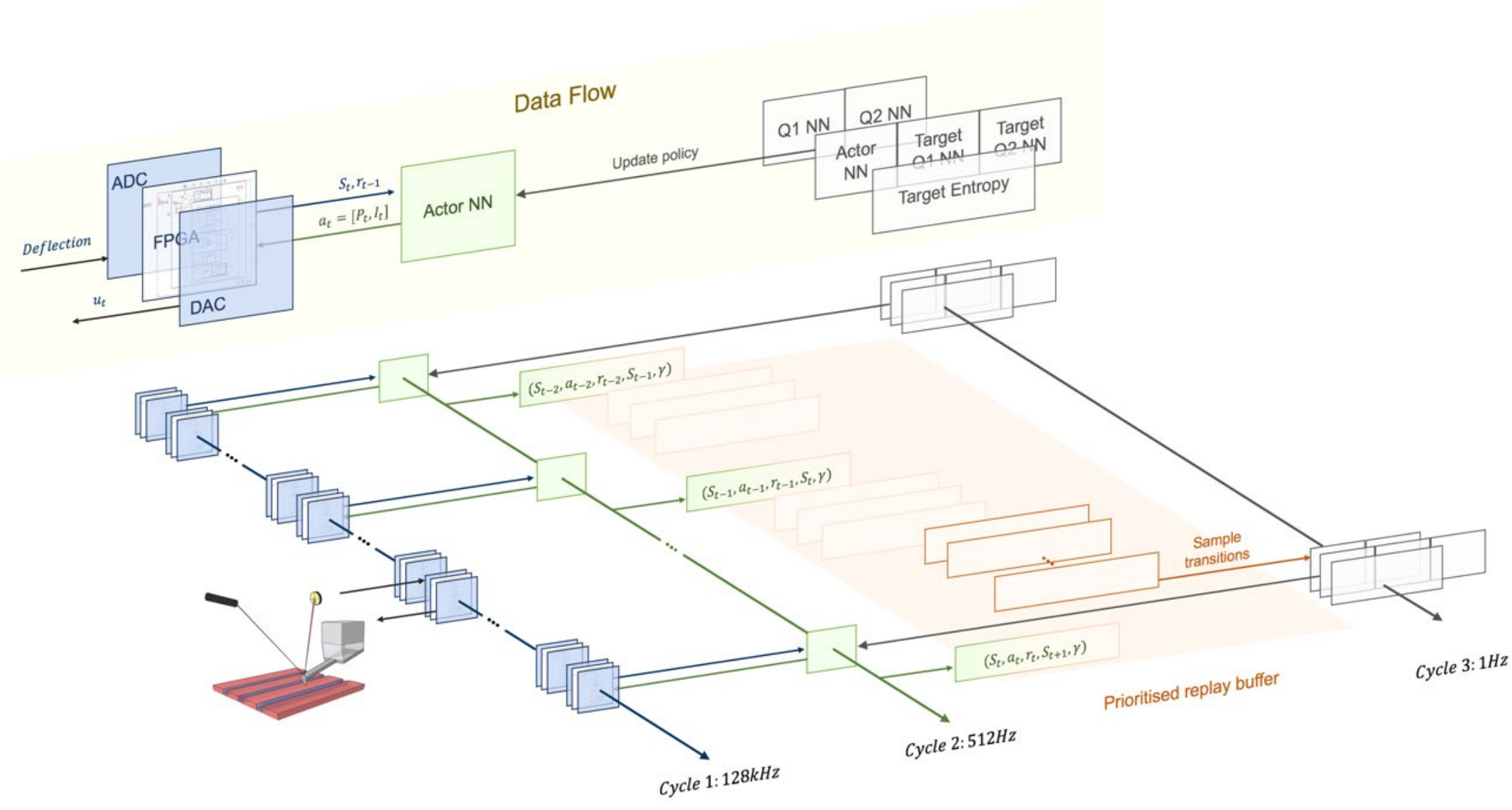

Figure 1, System architecture and data flow within the PICTS developed in this study.

The specific features of the PICTS are:

*Parallel Processing:* The three cycles operate concurrently but at different frequencies optimised for specific tasks, ensuring real-time responsiveness without overloading computational resources.

(a) Cycle 1 executes at the highest frequency (e.g., 128 kHz) on the FPGA, performing immediate and dynamic P-I control essential for rapid system response.
(b) Cycle 2 runs at a moderate frequency (e.g., 512 Hz) on the host computer, where the RL agent makes decisions to adjust control gains based on the current system state.
(c) Cycle 3 operates at a lower frequency (e.g., 1 Hz) for intensive training computations, updating the agent's policy to improve future decision-making.

*Adaptive Deep Reinforcement Learning Integration:* By abstracting the SPM control problem into a DRL framework, PICTS enables the agent to adjust control parameters dynamically. The system state $S_t$ at time $t$ includes:

(a) Error Signal ($e_t$): The deviation between the deflection voltage and the setpoint.
(b) Control Output ($u_t$): The control signal applied to the actuator.
(c) Historical Relative Positions ($H_t$): A sequence representing the cumulative control output over time, reflecting the actuator's historical movements. It provides the agent with temporal context, enabling it to understand system dynamics over time.
(d) Current P-I Gains ($P_t, I_t$): The P-I controller gains in use.

The agent's action space consists of incremental adjustments $a_t = [\Delta P_t, \Delta I_t]$ to the P-I gains, promoting smooth transitions and enhancing system stability.[21,24]

*2.1.3 Role of the Reward Function*

The reward function in DRL serves as a quantitative measure that guides an agent's behaviour by evaluating its actions within a given environment, driving the learning process toward desired outcomes. In this study, the design of the reward function is pivotal in shaping the agent's learning trajectory and ensuring robust control performance. As a critical component of the PICTS, the reward function addresses vital challenges:

*System Accuracy:* The primary objective is to reduce the error signal $e_t$ to ensure precise tracking and high-quality scanning results. In the initial learning phases (approximately the first ten scan lines during the scanning), the agent may select unstable combinations of P and I gains, leading to oscillations and abnormal fluctuations. To address this, we introduced a squared term into the reward function to smooth gradients and reduce instabilities. This squared term amplifies the penalty as the error increases, discouraging the agent from selecting unstable gain combinations and allowing the system to reach a stable state more quickly.

*Overshoot and Oscillation Reduction:* Overshoots can result in artificial peaks, while oscillations introduce noise, both can degrade image quality during the SPM scanning. The reward function incorporates penalties for overshoot and oscillations, guiding the agent to adjust the P-I gains to suppress these undesirable dynamics, the functions include following:

(a) *Overshoot Detection*: An FPGA-based Overshoot and Oscillation Detector (OOD) monitors the real-time error signal. When the error exceeds a defined threshold, the OOD enters a detection period and employs zero-crossing detection to confirm overshoots, assigning a penalty proportional to the squared error.

(b) *Oscillation Detection*: The OOD calculates the ratio of control steps where the error exceeds an oscillation threshold between two transitions. A penalty based on the square of this ratio is applied during training, hence discouraging oscillatory behaviour.

*Probe-Sample Contact Integrity:* Maintaining continuous contact between the probe and the sample is crucial for contact-based SPM techniques. The cantilever deflects asymmetrically around the setpoint, with unlimited deformation as the probe approaches to the sample until fracture and reduced deformation as it retracts. This asymmetry poses challenges, especially with low setpoint values (e.g., 0.1 V), which increases the risk of probe disengagement. The reward function includes a mechanism that penalises the potential disengagement, encouraging the agent to adjust its strategy to maintain stable contact between the probe and sample, resulting in high-quality results of SPM scanning.

By integrating these components into the reward function, the DRL agent is driven to optimise control actions that achieve precise tracking, minimise overshoot and oscillations, and ensure continuous probe-sample interaction. This results in improved control performance and higher-quality scanning outcomes.

## 2.2 Performance Evaluation

This section provides a detailed analysis of the scanning performance of the DRL-based controller integrated into the PICTS system, compared to the commercial controller in the SPM system. The results are organised into two main parts: qualitative case studies and quantitative statistical analyses. During the experiments, the commercial controller is implemented first to ensure that any distortions in its scanning results are not mistakenly attributed to potential sample damage caused by subsequent DRL-controlled scans. In this study, three materials are tested: (i) the SPM calibration grating, (ii) a biphasic polystyrene and low-density polyethene (PS-LDPE) sample, and (iii) a highly ordered pyrolytic graphite (HOPG). Those materials and tests are selected to demonstrate the effectiveness of the PICTS system and its integrated DRL controller in the challenging environments characterised by (i) **sharp edges (SPM calibration grating), (ii) soft multiphase structures (PS-LDPE sample), and (iii) significant topographical variations (HOPG sample)**, respectively. The PS-LDPE sample is specifically chosen because it represents a problematic test case for the normal contact mode SPM, which tends to generate shear forces that can damage soft materials and distort features.[1] In most studies, researchers tend to scan such samples using tapping mode to avoid these issues.[28,29] However, we employed contact mode on this traditional challenge material to highlight the advantages of the DRL controller developed in this study. Remarkably, the DRL-based controller performed exceptionally well, accurately capturing the morphology of the soft multiphase structure while mitigating oscillations and reducing artefacts and noise.

### *2.2.1 Qualitative Case Studies*

*(i) Sharp Edges sample (SPM calibration grating):* ***The SPM calibration grating is a 180nm deep, 10μm pitch platinum-coated grating mounted on a glass slide designed for precise 3D calibration of SPM (Bruker, SG, Asia Pacific).*** Fig. 2 (A1-A3) presents topography images of the calibration grating sample obtained using different control strategies. Subfigure A1 shows the results from the DRL controller, while A2 and A3 display outcomes from the commercial SPM controller with different integral gain settings. Subfigure B1-B4 analyses the scan parameters, including height, deflection signal, and P-I gains extracted from the horizontal line indicated in Subfigure A1. Subfigure C1-C4 shows the performance from the moderate-speed scanning, comparing the results from both controllers.

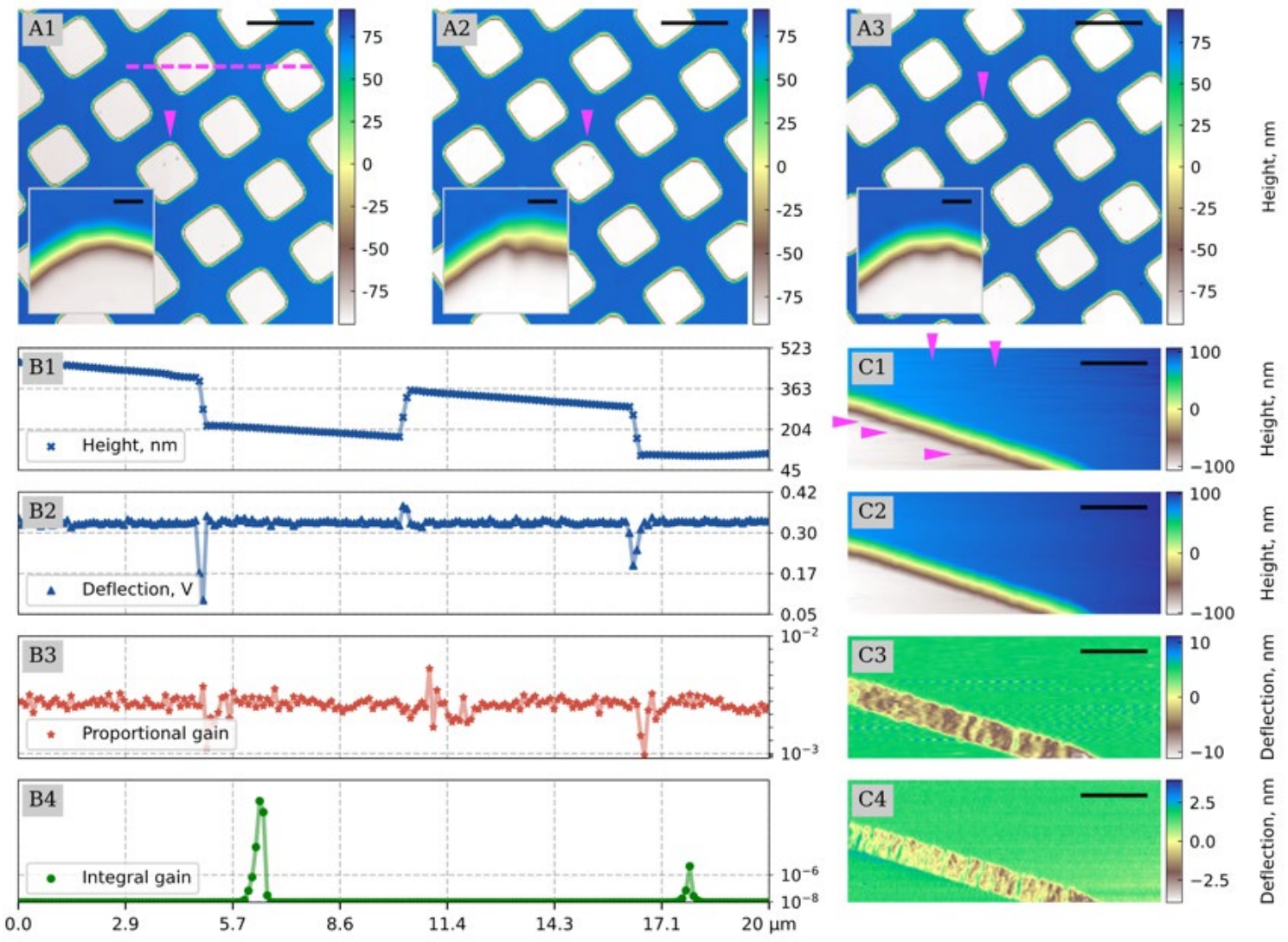

Figure 2, Performance comparison of the commercial SPM and DRL controllers on SPM Calibration grating sample. Subfigures A1-A3 presents 40×40 µm$^2$ topography images of the grating sample obtained using different control strategies under the control of PICTS, commercial SPM controller with integral I gain of 10, commercial SPM controller with integral I gain of 20 separately; scale bar for zoom-in area (indicated by red arrow in A1-A3): 260nm. Subfigures B1-B4 analyses the scan parameters—height, deflection signal, adjusted P gain, and I gain—extracted from the horizontal line area indicated in A1. Subfigures C1-C4 show results by moderate-speed (4Hz) scanning by using both controllers. C1 and C3 are the height and deflection signals by commercial SPM controllers, and C2 and C4 are those from the PICTS; scale bar: 1 µm.

**Comparative Analysis:** In Subfigure A1, the DRL controller captures the sample's topography over a 40 µm scan range, accurately tracking corners and sharp edges without distortions. By dynamically adjusting the P-I gains in real time, the DRL controller adapts to complex surface features and maintains precise adherence to the sample's morphology. In contrast, Subfigures A2 and A3 illustrate the limitations of a commercial SPM controller with fixed P-I parameters. Although an experienced operator initially sets the P-I gains, they remain constant throughout the whole scanning process. Consequently, the controller struggles to track complex surfaces accurately, leading to noticeable distortions, particularly at corners and sharp edges. In subfigure A2, where a lower gain setting (10, commonly used) is applied, distortions and protrusions are visible in the corner regions, indicating inadequate surface tracking due to insufficient gain. To improve monitoring, the gain was increased to 20 in the subfigure of A3. Although this higher gain setting enhances tracking compared to that of the subfigure A2, distortions persist around sharp edges. Additionally, the increased gain introduces gain noise into the system (Section S2 and Figure S2, SI), manifesting as unwanted fluctuations in the control signal that further degrade image quality. These observations highlight that, regardless of gain adjustments, the controller with fixed P-I parameters cannot fully eliminate topographic distortions and oscillations, underscoring its limitations in managing significant topographical variations.

**Dynamic Gain Adjustment with the DRL Controller:** Subfigures B1-B4 illustrate the DRL controller's ability to adjust P and I gains dynamically to adapt to complex sample topographies. To minimise transient errors, the P gain responds sharply to sudden height variations, such as edges. However, its adjustments differ significantly between upward and downward slopes: during upward slopes, where the probe maintains "passive contact" with the surface, the P gain changes are smaller as the natural contact conditions are inherently more stable. In contrast, on downward slopes, the P gain undergoes more dramatic adjustments to prevent the probe from losing contact. Once past these features, the P gain decreases to prevent overcompensation. Similarly, the I gain reflects this context-aware behaviour: it remains stable during upward slopes, as steady-state errors are minimal, but adjusts gradually after descending slopes to correct residual errors and ensure smooth probe motion. This nuanced response demonstrates the DRL controller's ability to intelligently interpret topographical features and maintain robust, precise tracking.

**Moderate-Speed Scanning Comparison:** To assess the adaptability of the DRL controller to scanning speed, **subfigure C1-C4** presents a comparison under moderate-speed scanning conditions (4 Hz), in which is the relevant high-speed for the SPM system used in this study (MFP-3D, Oxford Instruments, USA). The commercial SPM controller (Subfigure C1) fails to track fine topographic details accurately during the scanning speed of 4 Hz, resulting in significant stripe-like noise. In contrast, the DRL controller (Subfigure C2) maintains precise and stable topographic tracking with markedly reduced noise. Subfigures C3 and C4 show the deflection signals obtained with the commercial SPM and DRL controllers, respectively. In subfigure C3, the commercial controller’s deflection signal exhibits pronounced oscillations and a large amplitude, leading to the stripe-like noise observed in subfigure C1. Conversely, the deflection signal from the DRL controller remains stable, as shown in subfigure C4, with minimal error and almost no oscillations, contributing to the noise-free topography observed in subfigure C2.

*(ii) Soft Biphasic Structures:* ***The PS-LDPE is a polymer film with two distinction components; it features flatter polystyrene (PS) regions having an elastic modulus of around 2 GPa and spherical regions (LDPS) with an elastic modulus of around 0.1 GPa (Bruker, SG, Asia Pacific).*** Fig. 3 shows the trace topography, retrace topography and deflection images of the PS-LDPE sample scanned using the commercial SPM (subfigures A1-A3) and DRL (subfigures B1-B3) controllers. Subfigures C1-C4 illustrate the scanning parameters of the highlighted horizontal line in Subfigure B2 to show the DRL controller's adaptability at the biphasic interface.

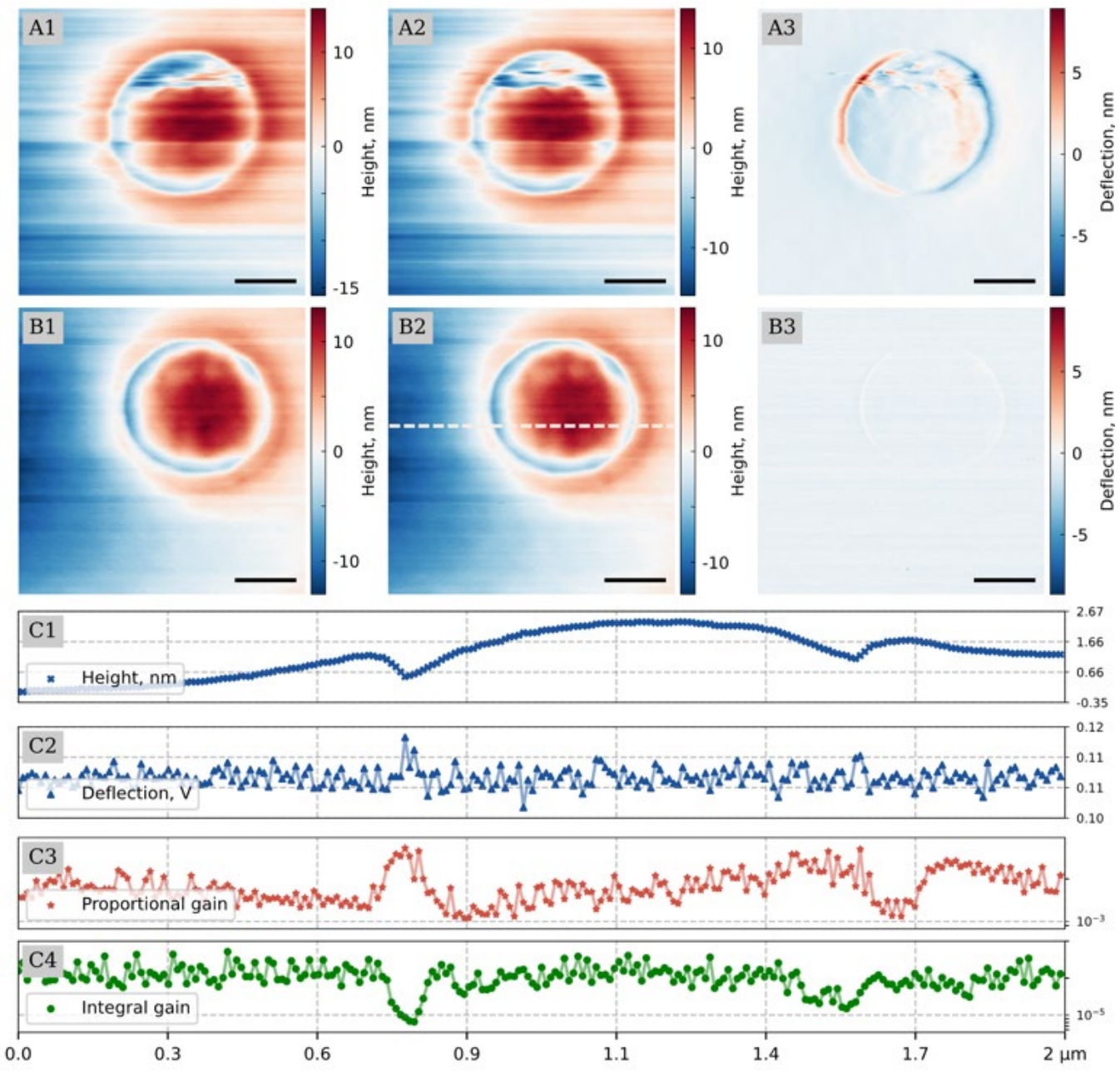

Figure 3, Performance comparison of the results by commercial SPM (subfigure A1-A3) and DRL (subfigure B1-B3) controllers on biphasic LDPE-PS sample. Scanning results from left to right: height trace, height retrace, and deflection signal for both controllers, respectively. Analysis of the highlighted horizontal line in B2, showing height, deflection, P gain, and I gain, are shown in the subfigures C1-C4. Scale bar: 400 nm.

**Comparative Analysis:** Comparing the subfigures A1-A3 and B1-B3 in Figure 3, the images from the commercial SPM controller (A1-A3) exhibit significant misalignment in the spherical LDPE region, with noticeable blurring and trailing artefacts at the edges. Despite minimal topographical variation at the biphasic interface, these artefacts primarily result from the substantial stiffness difference between the LDPE and PS regions (approximately 20-fold). When the probe transitions across the biphasic interfaces, the controller with fixed P-I parameters fails to adapt to the abrupt shift in the system state, causing the probe to either detach from or make hard contact with the material. As detailed in Section S3 of SI, the system state change is driven by sudden load disturbances and the nonlinear behaviour of the softer LDPE material, which destabilises the feedback loop. The controller lag effect further amplifies this instability by delaying the system's response to the sudden state change, leading to excessive force applied to the LDPE region. The low stiffness of LDPE makes it highly susceptible to deformation under these conditions, resulting in dragging effects and inaccuracies in topography measurement. In contrast, the DRL controller dynamically adjusts gain parameters in real time, allowing it to respond effectively to the changing system state at material transitions (subfigure B1 – B3). The deflection signal (B3) closely resembles the ideal condition, containing only minimal circuit noise. The DRL controller effectively eliminates deformation and trailing artefacts by maintaining stable contact, producing high-quality topography images with high trace-retrace overlap and minimal noise (B1 and B2). Additionally, the DRL controller successfully removes stripe noise across the entire image, further enhancing image quality.

**Dynamic Gain Adjustment with the DRL Controller:** subfigures C1-C4 indicate that the DRL controller employs a strategy significantly different from that used for the SPM calibration grating sample (Fig.2), demonstrating its ability to adapt to different probe-sample interactions. Significant gain adjustments primarily occur at the PS-LDPE interfaces of the sample. When transitioning from the stiffer PS to the softer LDPE, the P gain increases while the I gain decreases. This adjustment is necessary because the pliable LDPE requires a higher P gain to enable quicker responsiveness to deformation and a lower I gain to prevent excessive system oscillations. Conversely, when transitioning from the more deformable LDPE to the stiffer PS, a higher I gain is needed to eliminate persistent errors gradually, while a lower P gain is applied to prevent over-excitation of the rigid material.

*(iii) Sample with Significant Topographical Variations:* ***The HOPG sample features a freshly cleaved surface created by tape exfoliation, providing irregular height fluctuations of 4-6 microns (MikroMasch, SEA)***. Fig. 4 compares scanning results on the HOPG sample, highlighting the differences between the results obtained using the DRL controller and the commercial SPM controller, focusing on two distinct regions. The images are arranged in a matrix of four times four rows and columns, showing trace and retrace topography and deflection signals. The first two rows (subfigures A1-A8) depict the results for Region 1, scanned by the commercial SPM and DRL controllers, respectively. The last two rows (subfigures B1-B8) show the performance in Region 2 after four hours of continuous scanning, ensuring the system had reached to a steady state. Retrace images are flipped to align with the trace direction. More results can be found in Section S4 and Figures S4 and S5 of the SI.

**Comparative Analysis:** The commercial SPM controller, operating with fixed P-I parameters, struggled to handle the significant height variation of about 5.5 μm in Region 1. Ideally, the trace and retrace images should be symmetrical, representing consistent scanning of the same area in opposite directions. However, the commercial controller (subfigures A1-A4) results showed substantial discrepancies, with the trace and retrace scans almost unrecognisable as the same region, marred by artefacts and deflection errors. In contrast, the DRL controller's adaptability allowed it to adjust to large height changes, maintain stable tracking, and significantly reduce artefacts, producing more consistent and accurate trace and retrace images (subfigures A5-A8). After prolonged scanning, the commercial SPM controller exhibited stability improvements. While the trace-retrace overlap increased marginally, the controller struggled in regions with complex geometries, where it failed to track abrupt height changes accurately, leading to persistent artefacts (subfigures B1-B4). In contrast, the DRL controller maintained excellent stability throughout the extended scans and reduced deflection error to one-tenth of those observed with the commercial controller, ensuring sharper edges and better definition, significantly improving the trace-retrace alignment across the entire sample (subfigures B5-B8).

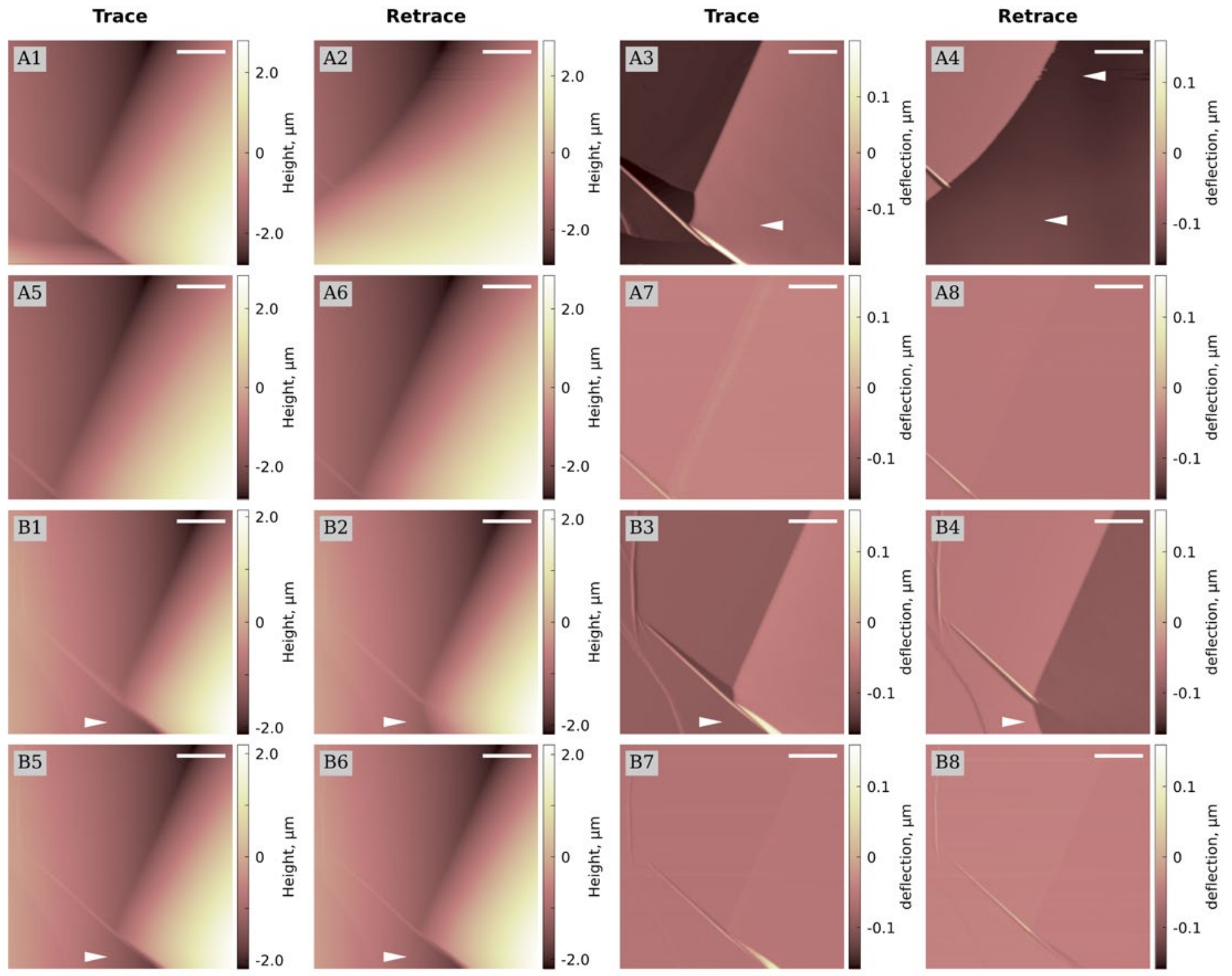

Fig. 4, SPM scanning results of 50×50 µm$^2$ HOPG sample surface using commercial SPM and DRL controllers across two regions. The images are arranged in 4x4 rows and columns, showing trace and retrace topography and deflection signals. The first two rows (subfigures A1-A8) depict the results for Region 1, scanned by the commercial SPM and DRL controllers, respectively. The last two rows (subfigures B1-B8) show the performance in Region 2 after four hours of continuous scanning, ensuring the system had reached a steady state. Retrace images are flipped to align with the trace direction. Scale bar: 10 µm.

### *2.2.2 Quantitative Statistical Analyses*

Fig. 5 presents a statistical comparison between the results from the commercial SPM and DRL controllers across various samples, scanning speeds, and scanning ranges. The results include ten histograms illustrating the error distribution for each scenario (Fig. 5A) and an error magnitude plot (Fig. 5B). Detailed scanning results, i.e., height and deflection images for various samples, including two additional samples (aerogel and carbon fibre reinforced polymer, CFRP), details and analysis can be found in the Sections S5 and S6, Figures S6 to S8 of the SI.

Fig. 5A presents histograms of the error distribution for each scenario. The DRL controller consistently exhibits a higher concentration of error values near zero, indicating greater precision and fewer outliers. Furthermore, extreme errors are significantly lower with the DRL controller, highlighting its robustness in maintaining consistent performance, even under challenging scanning environments.

Fig. 5B displays the mean magnitude of the deflection error, i.e., the discrepancy between the setpoint voltage and the deflection voltage, for different samples. Compared to the commercial SPM controller, the DRL controller achieves consistently lower mean deflection errors, with reductions ranging from 26% (grating sample) to as much as 90% (HOPG sample). Notably, the most significant reductions occur with samples with complex interfaces (e.g., PS-LDPE) and challenging topographies (e.g., HOPG with height variations 4-6 μm), demonstrating the superior adaptability of the DRL controller under demanding conditions.

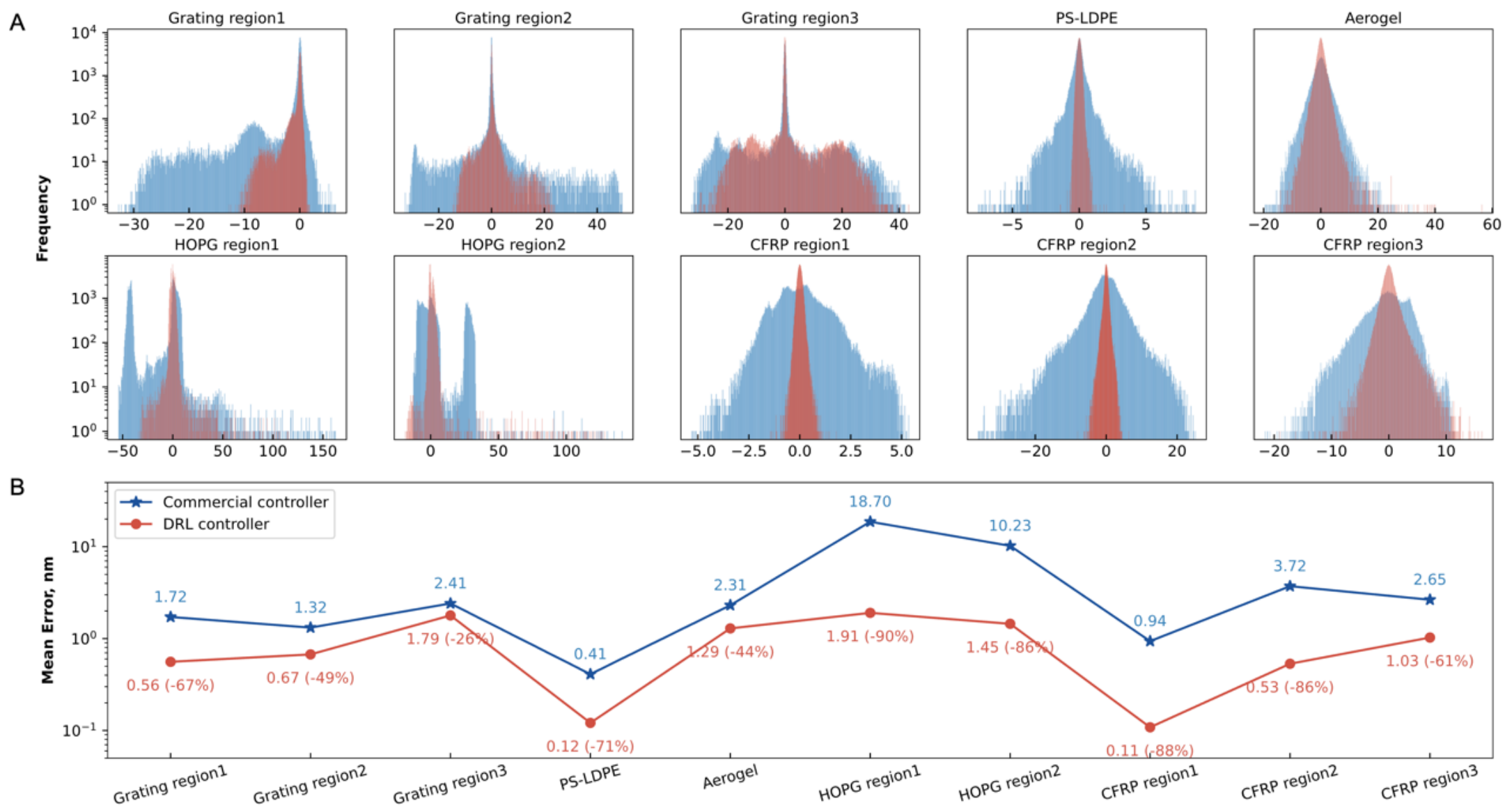

Fig. 5, Statistical comparison of magnitudes and distributions of the deflection error from the commercial SPM and DRL controllers across various samples and scanning conditions. The figure includes ten histograms (A) and an error magnitude plot (B) showing the distribution of deflection errors. The blue and red colours represent results from the commercial SPM and DRL controllers, respectively, across the following sample conditions: 5×5μm$^2$ at 4 Hz (SPM calibration grating), 20×20 μm$^2$ at 1 Hz (SPM calibration grating), 40×40 μm$^2$ at 1 Hz (SPM calibration grating), 2×2 μm$^2$ at 1 Hz (PS- LDPE), 5×5 μm$^2$ at 1 Hz (aerogel), 50×50 μm$^2$ at 1 Hz (HOPG regions 1 and 2), 2×2 μm$^2$ at 1 Hz (carbon fibre reinforced polymer or CFRP), and 10×10 μm$^2$ at 1 Hz (CFRP regions 2 and 3).

These quantitative and statistical results confirm that the DRL controller outperforms the commercial SPM controller across various samples and scanning conditions. By dynamically adjusting control gains in real time, the DRL controller enhances accuracy and stability, making it a valuable tool for improving SPM measurements.

## 3. Conclusion

Controlling mechanisms in Scanning Probe Microscopy (SPM) face significant challenges when operating in highly nonlinear and dynamic environments. Complex sample surfaces and various probe-sample interactions necessitate experts to tune controllers and achieve high-quality images, making it difficult to distinguish morphological features from artefacts consistently. To address these limitations, we have developed a novel control strategy integrating deep reinforcement learning (DRL) into tuning proportional-integral (P-I) controllers.

By framing the SPM control problem within a reinforcement learning framework, our DRL-based controller adaptively adjusts P-I gains in real time based on the system's state. Trained with a tailored reward function, the DRL agent effectively balances, minimising tracking errors, reducing overshoot and oscillations, and maintaining stable probe-sample contact. This adaptive approach enables the controller to handle sudden changes in sample properties and topographical features, significantly improving system stability and scanning precision.

The Parallel Integrated Control and Training System (PICTS) developed in this work can outperform the traditional controller with fixed P-I parameters, particularly when dealing with sharp edges, multiphase soft structures, and significant topographical variations. Statistical analyses demonstrate that the DRL controller reduces deflection errors by **26% to as much as 90%**, depending on the nature of the materials, with error values more concentrated around zero. These improvements highlight the controller's superior performance under challenging conditions.

Furthermore, the PICTS architecture efficiently balances real-time control and computational demands by leveraging an FPGA to handle critical tasks without overloading system resources. This design maintains high responsiveness and is accessible to laboratories with standard hardware, eliminating the need for expensive computational setups.

In summary, integrating DRL into SPM control systems offers a robust solution for complex and dynamic scanning environments, enhancing image quality and expanding the capabilities of SPM technology. Our work demonstrates the potential of advanced machine learning techniques to overcome the limitations of traditional control methods, paving the way for further innovations in the field.

## 4. Methods

### 4.1 Technical settings

In implementing the DRL control strategy for the SPM system, we create a lightweight system that can operate efficiently on the host computer, equipped with a standard CPU. The system architecture is designed to ensure efficient coordination between the FPGA, host computer, and the commercial controller, leveraging high-speed analogue/digital signal processing to meet the demanding requirements of SPM control. All contact-mode SPM measurements were conducted using the commercial SPM system (MFP-3D, Asylum Research, Oxford Instruments, CA, USA).

### *4.1.1 System Integration and Communication*

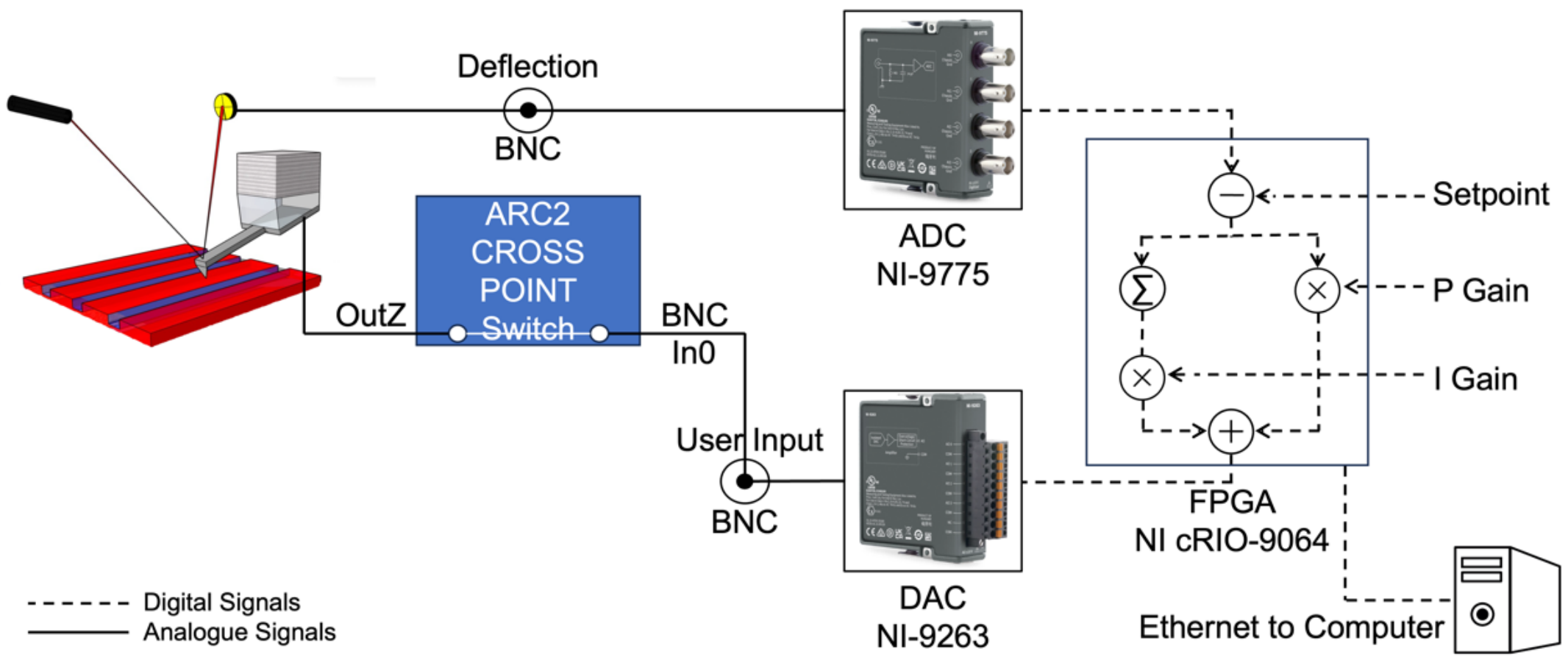

Fig. 6, Signal Integration and Data Flow in PICTS. Using FPGA and Analogue /Digital Signal Integration for SPM Control.

The control system developed in this work seamlessly integrates various hardware and software components:

*(i) Hardware Components consist of following:*

FPGA (**NI CompactRIO FPGA, cRIO-9064**): It executes the high-speed P-I control loop and computes specific components of the reward function.

ADC Module (**NI-9775**): It reads the deflection voltage, providing accurate input for the P-I controller.

DAC Module (**NI-9263**): It outputs control voltage to the Z-axis piezoelectric actuator for precise probe positioning.

**Low-Pass Filter**: This reduces signal noise on the analogue output to prevent interference-induced fluctuations.

*(ii) Software Components consist of following:*

**Host Computer**: It runs Python code for the RL agent (Cycle 2, Fig.1) and the training thread (Cycle 3, Fig.1).

**Soft Actor-Critic (SAC) Algorithm**: This facilitates efficient and stable training of the DRL agent.

**nifpga Python Library**: It enables efficient communication between the FPGA and the host computer for real-time updates.

*(iii) Communication Protocols:*

**Analogue Input-Output Communication**: This operates via TCP over Ethernet, with a maximum data transmission speed of 100 MB/s and a transmission delay of 1000 μs.

**Real-Time Data Exchange**: It ensures the DRL agent receives timely feedback for decision-making and that the FPGA receives updated control actions without significant latency.

By distributing tasks between the FPGA and the host computer in this parallel architecture, we address the limitations of control speed and precision inherent in standard systems. The FPGA manages real-time control tasks and partial reward computation, while the host computer handles decision-making and learning processes, optimising overall system performance.

*4.1.2 SPM settings*

*SPM Probe and Samples:* Standard calibration samples (VGRP-15M and PS-LDPE-12M) were obtained commercially (Bruker AFM Probes International, SG, Asia Pacific). The samples HOPG were purchased commercially (MikroMasch, Asia Pacific). These samples were used to validate the performance of the SPM system and assess the effectiveness of the DRL control strategy across various topographical features. Two types of probes were used in the experiments (240AC-NA from OPUS by MikroMasch and PPP-CONTR from NANOSENSORS).

*Performance Considerations:* The system design in this work aligns with the operational needs of SPM control. When scanning for high-quality SPM imaging, the P and I values must adjust according to the topography but should not change so frequently to cause undesirable oscillations. Therefore, we set the adjustment frequency of the P and I gains to match the XY spatial resolution during the SPM imaging, that is, one adjustment per scan point. For instance, at a scan rate of 1 Hz for an SPM image with any scan range and a resolution of $256 \times 256$ pixels, Cycle 2 requires a running rate of 512 Hz. To maintain precise control, Cycle 1 necessitates a sampling rate of 128 kHz. The parallel architecture of the designed controller supports these rates, allowing for rapid, real-time control adjustments while retaining the computational efficiency necessary for practical laboratory applications.

## 4.2 Deep Reinforcement Learning

*4.2.1 SAC Network Architecture*

The training strategy employs the SAC algorithm with the Automating Entropy Adjustment method. SAC is an off-policy reinforcement learning algorithm designed for continuous action spaces, balancing exploration and exploitation by maximising a trade-off between expected return and policy entropy. The SAC implementation involves five (5) neural networks: one (1) **Policy Network (Actor)**, which outputs continuous actions to adjust the P-I gains of the controller based on the current state of the system; two (2) **Q-Value Networks (Critics)**, which estimate the expected return (Q-values) for state-action pairs to evaluate the policy; and two (2) **Target Q-Value Networks**, which serve as stable targets for the critics, updated using a soft update mechanism.

### *4.2.2 Training Details*

Due to the parallel architecture of the system, training runs continuously alongside the data collection during the scanning of images. The agent samples experiences from a prioritised replay buffer,[30] which improves learning efficiency by focusing on more informative transitions. **Prioritisation Exponent (α)** set to 0.5, balancing between uniform sampling ($\alpha=0$) and full prioritisation ($\alpha=1$). This parameter controls how much prioritisation is used in the sampling; higher values of α increase the preference for transitions with higher temporal-difference (TD) errors. **Importance Sampling Weight Exponent (β)** set to 0.5, it is used to correct the bias introduced by prioritised sampling. It adjusts the importance-sampling weights to ensure that the updates remain unbiased in expectation. The experience replay buffer stores up to 1,000,000 transitions, providing a rich dataset for training. The agent samples batches of size 256 for each training update.

We utilise the "**Adam**" optimiser for all network updates, which is known for its efficiency and adaptive learning rates. To further enhance convergence and prevent overfitting, we integrated two learning rate schedulers: **Cosine Annealing Learning Rate Scheduler (CosineAnnealingLR)**, which gradually reduces the learning rate following a cosine curve without abrupt changes, helping the optimiser to fine-tune the policy as training progresses, and **Reduce on Plateau Scheduler (ReduceLROnPlateau)**, which monitors the training performance and reduces the learning rate when a metric has stopped improving, allowing the optimiser to escape potential local minima.

In this work, all hyperparameters and configurations are determined through extensive testing and grid searches in **a simulated environment**. Using prioritised experience replay and automatic entropy adjustment contributed to stable learning and efficient exploration.

## Acknowledgement

This research was supported by the Ministry of Education (MOE) Singapore through the National University of Singapore (NUS) under the Academic Research Funds (Nos. A-0009122-01-00). The author would like to thank the financial support from the NUS Research Scholarship (Industry Relevant) (RO-Research Scholarship IR FOE IS and GRSUR0600039). We would like to thank Mr. Guiyin Xu from and Dr. Xinyu Dong from the Department of Mechanical Engineering, NUS, for generously providing the CFRP and Aerogel samples for this study.

## Supplementary Information

The Supplementary Information (SI) of this paper can be obtained upon requirements.